\documentclass[showpacs,twocolumn,aps,prl,superscriptaddress]{revtex4}
\usepackage{amsmath,amssymb}
\usepackage{graphicx}
\usepackage{dcolumn}
\usepackage{color}
\usepackage{epsfig}
\usepackage{epstopdf}
\usepackage{subfigure}
\usepackage{times}
\usepackage[italian]{babel}
\usepackage{bbold}
\usepackage{amsfonts}
\usepackage{dsfont}
\newcommand{\sys}{\cal S}
\newcommand{\enm}{{\cal E}_{nM}}
\newcommand{\ema}{{\cal E}_{M}}

\DeclareMathSizes{10.5}{10.5}{1}{1}

\usepackage{bm}

\newcommand{\ket}[1]{\left\vert#1\right\rangle}

\graphicspath{./figs/}

\begin{document}

\title{Competition between memory-keeping and memory-erasing decoherence channels}

\author{Tony~J.~G.~Apollaro}
\affiliation{Dip.  Fisica, Universit\`a della Calabria, 87036
Arcavacata di Rende (CS), Italy} \affiliation{INFN - Gruppo
collegato di Cosenza, Cosenza Italy} \affiliation{Centre for
Theoretical Atomic, Molecular and Optical Physics, School of
Mathematics and Physics, Queen's University, Belfast BT7 1NN,
United Kingdom}

\author{Salvatore Lorenzo}
\affiliation{Dipartimento di Fisica e Chimica, Università\`{a}
degli Studi di Palermo, via Archirafi 36, I-90123 Palermo, Italy}

\author{Carlo Di Franco}
\affiliation{Centre for Theoretical Atomic, Molecular and Optical
Physics, School of Mathematics and Physics, Queen's University,
Belfast BT7 1NN, United Kingdom}

\author{Francesco Plastina}
\affiliation{Dip.  Fisica, Universit\`a della Calabria, 87036
Arcavacata di Rende (CS), Italy} \affiliation{INFN - Gruppo
collegato di Cosenza, Cosenza Italy}

\author{Mauro Paternostro}
\affiliation{Centre for Theoretical Atomic, Molecular and Optical
Physics, School of Mathematics and Physics, Queen's University,
Belfast BT7 1NN, United Kingdom}

\pacs{}

\begin{abstract}

We study the competing effects of simultaneous Markovian and
non-Markovian decoherence mechanisms acting on a single spin. We
show the existence of a threshold in the relative strength of such
mechanisms above which the spin dynamics becomes fully Markovian,
as revealed by the use of several non-Markovianity measures. We
identify a measure-dependent nested structure of such thresholds,
hinting at a causality relationship amongst the various
non-Markovianity witnesses used in our analysis. Our
considerations are then used to argue the unavoidably
non-Markovian evolution of a single-electron quantum dot exposed
to both intrinsic and Markovian technical noise, the latter of
arbitrary strength.

\end{abstract}
\maketitle


The dynamics of open quantum systems~\cite{BreuerPbook} is of
considerable interest both from a fundamental perspective, e.g.
for the study of the quantum-to-classical
transition~\cite{Joosetalbook}, and for quantum technology. In the latter context,
it is important to understand and characterize the effects that an
environment has on the quantum features of a given
system~\cite{NielsenCbook}. 
One can distinguish between non-Markovian dynamics, where the
coupling with the environment results in the revival of the
coherences in the state of the system, and Markovian ones for
which no environmental back-action occurs. In many realistic
situations, system $\mathcal{S}$ interacts (at different
characteristic timescales) with distinct physical environments.
This is the case, for instance, for an electron in a quantum dot
(QD) that is strongly coupled to the surrounding nuclei and, more
weakly, to the phonons of the substrate where it has
grown~\cite{HansonKPTV07}, for a nitrogen-vacancy (NV) center
coupled to a bath of spins, embodied by the nitrogen impurities in
diamond, where the carbon-13 nuclear spins, on the other hand,
couple only weakly to the NV spin~\cite{Hansonetal08}, and for a
single-donor electron spin in
silicon~\cite{Plaeetal12,Tyryetal12}, where, thanks to the
availability of a isotopically enriched ${}^{28}$Si form without
magnetic nuclei, and in presence of a magnetic field gradient,
interactions with other donors (with non-zero nuclear magnetic
moment) are almost suppressed.

While even the intuitive picture of a non-Markovian process
provided above is not universally agreed on and we lack of a
general consensus on the very meaning of memory--keeping dynamics
in the quantum realm, some theoretical tools have been recently
proposed and used to characterize the degree of the non-Markovian
nature in the dynamics of open quantum systems. Measures aiming at
quantifying any deviation from a Markovian evolution have been put
forward~\cite{BreuerLP09, LorenzoPP13, RivasHP10, LuoFS12, bogna}
and applied to a number of physical situations~\cite{models},
including the first experiments pointing towards the controlled
simulation of non-Markovian dynamics~\cite{experiments}.

Notwithstanding a few attempts aimed at finding a unification and
an ordering for such a variety of
tools~\cite{BreuerLP09,Chruciski1KR11,HaikkaCM11}, and apart from
the simple case in which only a {\it single} decoherence channel
is present \cite{uncanale}, where all of the previous proposals
are essentially equivalent, a conclusive picture has yet to be
found. In this work we perform some significant steps in this
direction by considering the effects of simultaneous environmental
mechanisms. Our aim is to investigate open-system dynamics in the
presence of {\textit{competing}} effects arising because of the
simultaneous presence of various channels on a system of interest.
We will consider the interaction between the system and an
environment ${\mathcal{E}}_{nM}$ that induces strong non-Markovian
features, and, at the same time, will assume the system to be also
exposed  to the influences of a completely forgetful channel
${\mathcal{E}_M}$ that, on its own, would be responsible for
Markovian evolution. Under these conditions, we determine the
amount of a ``standard white noise'' (enforcing a Markovian
behavior) that one needs to add to the given memory--keeping
channel in order to let it become fully memoryless.

We show that it is possible to identify precise working conditions
under which the system evolution changes from Markovian
to non-Makovian and viceversa, thus highlighting the way channels
of different nature mutually interfere so as to induce a radical
change in the character of a dynamics. This is becoming
increasingly important in light of schemes that have been recently
proposed for quantum state engineering and quantum control based
on non-Markovian evolutions~\cite{control}.

Before entering the core of our analysis, it is beneficial to
review shortly the features of the various tools for the analysis
of non-Markovianity that have been mentioned above. A common
aspect  of such quantifiers is that they are expressed in terms of
non-Markovianity rates $X_l(t)$ (also termed {\it ${\cal
N}_l$-rates} in the following, with label $l$ used to distinguish
the various cases that will be addressed below). For Markovian
dynamics, these quantities are supposed to stay negative (or
constant) in time, while any deviation from negativity signals the
non-Markovian nature of an evolution, which is quantified by
integrating over the time windows where $X_l(t)>0$
\begin{equation}\label{E.definitionsnm}
\mathcal{N}_l{=}\int_{X_l(t){>}0}X_l(t)~{dt}.
\end{equation}
A few choices for $X_l$ have been made so far~\cite{BreuerLP09,
LorenzoPP13, RivasHP10, LuoFS12}, implying that different {\it
evidences} of non-Markovianity can be  gathered from the very same
dynamics: The measure proposed in Ref.~\cite{BreuerLP09}, denoted here
as ${\cal N}_{BLP}$, is based on the observation that a
non-Markovian evolution induces a non-increasing behavior of the state distinguishability.
In Ref.~\cite{LorenzoPP13}, some of us proposed to consider
$X_{LPP}{=}\partial_t{\left|\left|{F}\right|\right|}$, where ${F}$
is the affine transformation of the Bloch vector of the system
induced by a dynamical map and $\left|\left| {F}\right|\right|$ is
the absolute value of its determinant. The corresponding measure
${\cal N}_{LPP}$ is based on the assumption that a Markovian
dynamics induces a monotonic shrinking of the volume of accessible
states of $\mathcal{S}$. According to Ref.~\cite{RivasHP10},
Markovianity is equivalent to the divisibility  of the dynamical
map $\Phi_{(t,0)}$ (defined so that $\rho^{\sys}_t=\Phi_{(t,0)}
\rho^{\sys}_0$, where $\rho^{\sys}_t$ is the state of the system
at time $t$).

Writing $\Phi_{(t+\tau,0)}=\Phi_{(t{+}\tau,\tau)}\circ\Phi_{(t,0)}$ with
$\circ$ standing for the composition of two maps, Markovianity is
mathematically translated into the condition that
$\Phi_{(t+\tau,t)}$ should be a completely positive and trace
preserving (CPT) map for all $t$ and $\tau$. This implies that one
can relate the integrand of ${\cal N}_{RHP}$ in
Eq.~\eqref{E.definitionsnm} to the deviation from positivity of
the Choi matrix (isomorphic to the map $\Phi_{t{+}\tau,\tau}$).
Finally, in Ref.~\cite{LuoFS12} Markovianity is
synonymous of monotonic decrease of the mutual information
between the system and an ancilla whose joint initial state is
maximally entangled.

Let us now consider a system ${\mathcal{S}}$
interacting with two environments, ${\mathcal{E}_M}$ and
${\mathcal{E}}_{nM}$ under the assumption that, if ${\mathcal{S}}$
interacted only with the former (latter), its dynamics would be
Markovian (non-Markovian). In order to fix the ideas, we consider
both ${\cal S}$ and ${\cal E}_{nM}$ embodied by a spin-${1}/{2}$
particle. The action of the environment $\ema$ on the state $\rho_t$ of the
${\cal S}$-${\cal E}_{nM}$ system is described by the master
equation
$\dot{\rho}_t{=}{-}i[{\hat{\cal{H}}},{\rho_t}]+{\mathcal{L}}_t\rho_t$,
where $\hat{\cal H}$ is the Hamiltonian describing the ${\cal
S}$-$\enm$ dynamics and ${\cal L}_t$ is the Liouville
super-operator
\begin{equation}\label{E.Lindblad}
{\mathcal{L}}_t\rho_t{=}\sum_{k,j{=}1}^3\!\!\frac
{\gamma_{kj}}{2}\left([\hat\sigma^{\sys}_k,\rho_t{{\hat\sigma}^{{\sys}\dagger}_j}]{+}
[{\hat\sigma}^{\sys}_k\rho_t,{{\hat\sigma}^{{\sys}\dagger}_j}]\right)
\end{equation}
that describes the Markovian dynamics that would be enforced by
$\ema$ only. Here, $\hat\sigma^{\sys}_j$ is the $j$-Pauli matrix
of $\sys$~($j=x,y,z$), and $\gamma_{kj}$ are the entries of the
(hermitian) Kossakowski matrix ${\bm \gamma}$~\cite{GKS,Daffer}.
The dynamical map corresponding to the master equation above is
completely positive for ${\bm\gamma}\ge0$. In what follows, we
will restrict our attention to Kossakowski matrices taking the
form ${\bm \gamma}{=}{{\left(
\begin{matrix}
\gamma_x  & \alpha{+}i\beta & 0 \\
\alpha{-}i\beta & \gamma_y & 0 \\
0 & 0 & \gamma_z
\end{matrix}\right)}}$.
Although this choice does not embody the most general case, it
encompasses several relevant quantum channels~\cite{RomeroLF12}.
For instance, for $\alpha=\beta=0$ we have a Pauli
channel~\cite{cerf}, while an amplitude damping channel is
retrieved for $\alpha=\gamma_z=0$ with
$\beta=2\gamma_{x,y}{=}\gamma$~\cite{NielsenCbook}.

Finally, we take the $\sys$-$\enm$ interaction to be of the
Ising-type,
$\hat{\cal{H}}{=}({J}/{2})\hat\sigma_{A}^z\hat\sigma_{\enm}^z$
and, from now on, use the exchange constant $J$ as our frequency
unit. This choice allows us to provide an interesting analysis
without introducing unnecessary complications inherent in other
coupling models, and, in addition, it links to the experimental
scenarios we are going to address.

The reduced dynamics of $\sys$, described by the dynamical map
$\rho^{\sys}_t{=}\phi_t(\vec{\gamma},\alpha,\beta)\rho^{\sys}_0$
(with $\rho^{\sys}_t{=}{\rm Tr}_{\enm}[\rho_t]$) is completely
characterized by the choice of the entries of ${\bm\gamma}$, here
identified by the vector
$\vec{\gamma}{=}(\gamma_x,\gamma_y,\gamma_z)$ and by the
parameters $\alpha$ and $\beta$. As initial conditions, we
consider the factorized state
$\rho_0{=}\rho^{\sys}_0{\otimes}\rho^{\enm}_0$ with
$\rho^{\sys}_0{=} {{
\begin{pmatrix}
A^{++}_0&A^{+-}_0\\
A^{-+}_0&A^{--}_0
\end{pmatrix}}}$
and
$\rho^{\enm}_0{=}
{{
\begin{pmatrix}
B^{++}_0&B^{+-}_0\\
B^{-+}_0&B^{--}_0\end{pmatrix}}}$, both written in the basis
$\{\ket{\pm}\}$ of eigenstates of $\hat\sigma_z$. Due to the form
of the coupling chosen above, a prominent role will be played by
the initial magnetization of $\enm$. We therefore introduce the
parameter $z{=}2B^{--}_0{-}1$. The evolved state of the system has
matrix elements
\begin{equation}
\begin{aligned}
&A^{++}_t=\sum_{p=\pm}A^{pp}_{0}\left(\frac{f_t^{p}}{2}-\frac{\beta f_t^{-}}{\gamma_x{+}\gamma_y}\right),\\
&A^{+-}_t=A^{+-}_0[\text{Ch}{+}i
 \, z \, \text{Sh}]{+}A^{-+}_0(\gamma_x{-}\gamma_y-2i
\alpha)\text{Sh}
\end{aligned}
\end{equation}
where we have introduced the functions $f_t^{\pm}{=}1{\pm}
e^{-2(\gamma_x{+}\gamma_y)t}$ that account for the dissipative
action of $\ema$, and the short-cut notation
$\lbrace\text{Ch},\text{Sh}\rbrace{=}e^{{-}(\gamma_x{+}\gamma_y{+}2\gamma_z)t}\lbrace\cosh\xi
t,\frac{\sinh\xi t}{\xi}\rbrace$ with
$\xi{=}\sqrt{(\gamma_x{-}\gamma_y)^2{+}4\alpha^2{-}1}$. These
embody both the decoherence induced by $\ema$ and the
(non-Markovian) dynamics due to the coupling with
$\mathcal{E}_{nM}$. The populations of $\rho^{\sys}_t$ do not
depend on the parameters of $\hat{\mathcal{H}}$ nor on
$\rho^{\enm}_0$ due to the fact that the $\sys$-$\enm$ coupling is
dissipationless. In addition, the coherences are symmetric under
the exchange of $B^{--}_0$ with $B^{++}_0$ (or $z$ with $- z$) and
viceversa.


With the dynamical map $\phi_t(\vec{\gamma},\alpha,\beta)$, we can
now evaluate some of the measures of non-Markovianity recalled
above. Let us start with the isotropic depolarizing channel acting
on $\sys$. This is set by taking
$\gamma_x{=}\gamma_y=\gamma_z=\gamma_0/4$ and ${\alpha,\beta}=0$.
In this case, we can provide compact analytical expressions for
the ${\cal N}$-rates $X_l$ at the core of
Eq.~\eqref{E.definitionsnm} for the four measures discussed above.
We find
\begin{equation}
\label{analitiche}
\begin{aligned}
X_{BLP}&=\partial_t|G_t|,\,\,\,\,\,X_{LPP}=\partial_t\left[(2f_t{-}1) |G_t|^2\right],\\
X_{RHP}&=\lim_{\tau \to 0^+}\frac{1}{2\tau}\left(g(t,\tau){-}f_\tau{+}
\left|g(t,\tau){-}f_\tau\right|\right),\\
X_{LFS}&{=}\partial_t[2{+}2h(1{-}f_t)+\sum_{s{=}\pm}h(f_t{+}s|G_t|)],
\end{aligned}
\end{equation}
where $h(x){=}(x/2)\log_2(x/2)$, $G_t{=}\sum_{p{=}\pm}e^{-\gamma_0
t{+} i p t} B^{pp}_0$, $f_t{=}{(1{+}e^{-\gamma_0 t})}/{2}$, and
$g(t,\tau){=}\left|{G_{t+\tau}}/{G_t}\right|$. The dynamics of
$\sys$ is non-Markovian for any pair $(\gamma_0, z)$ such that the
quantities in Eqs.~(\ref{analitiche}) are positive. For the
$N_{RHP}$ measure, this condition leads to
\begin{equation}
\mu(z,t,\gamma_0)\equiv\gamma_0{+}\frac{(1-z^2)\sin(2
t)}{1-(1-z^2)\sin^2 t}{<}0. \label{ineqRHP}\end{equation} As the
function $\mu(z,t,\gamma_0)$ is even in $z$, we can take
$z{\in}\left(0,1\right]$, and obtain
$\gamma_{RHP}^*{=}(1{-}z^2)/{z}$ as a threshold for the onset of
Markovianity.

This means that $\mathcal{N}_{RHP}{=}0$ and the dynamics is
Markovian for all $\gamma_0\geq\gamma_{RHP}^*$. The case of
$z{=}0$, associated with equal diagonal entries in
$\rho^{\enm}_0$, deserves special attention as in this case
$\gamma_{RHP}^*\to\infty$, implying that the reduced
dynamics  will always be non-Markovian, regardless of the
strength of the coupling between $\sys$ and $\ema$. Following a similar
procedure we find the thresholds for the BLP and LPP measures as
$\gamma_{BLP}^*{=}(3/2)\gamma_{LPP}^*=(1{-}z^2)/2 z$.

\begin{figure}
 \centering
 {\includegraphics[width=\columnwidth]{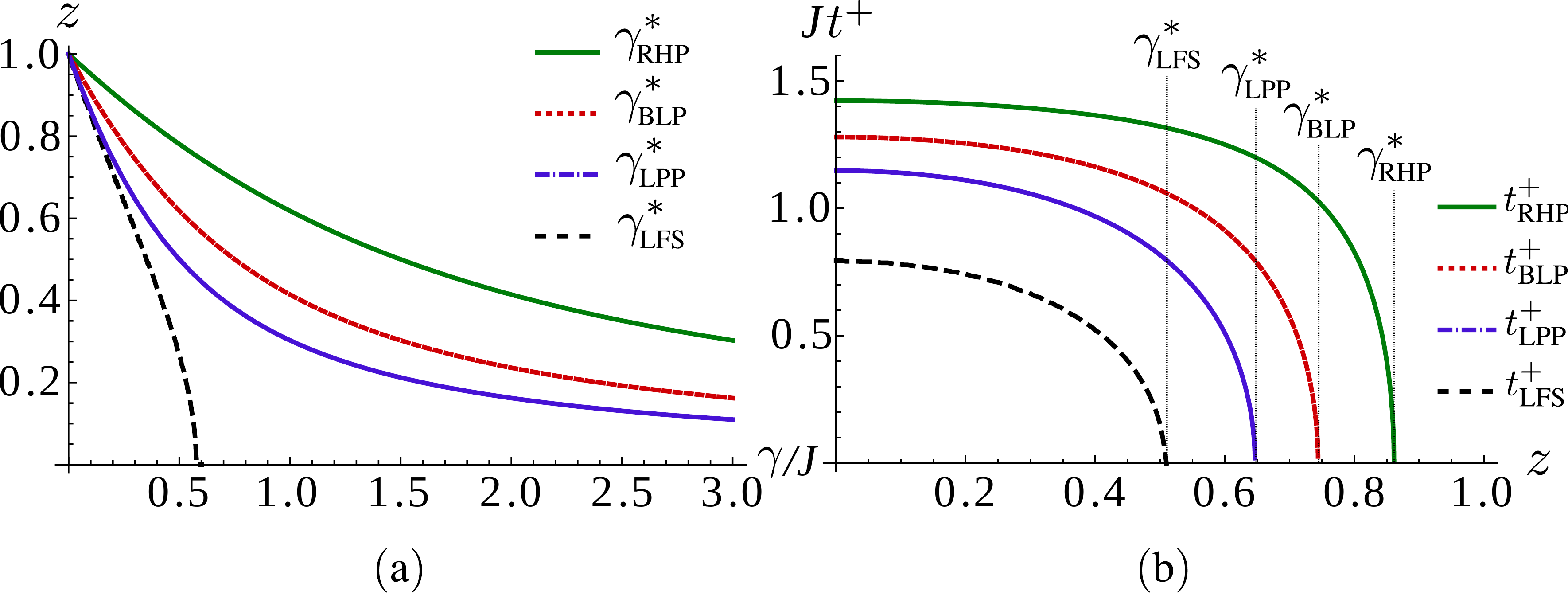}}
\caption{(Color online) {\bf (a)} Markovianity phase diagram in
the $(\gamma,z)$-plane. In the region above (below) each curve,
the dynamics of $\mathcal{S}$ is Markovian (non-Markovian)
according to the corresponding measure. {\bf (b)} Hierarchy of the
time windows giving contributions to Eq.~\eqref{E.definitionsnm}.
In the cases corresponding to the three upper curves, the time
evolution is periodic, and positivity intervals recur in time. For
the LPS measure, due to lack of periodicity, we only report the
first positivity interval, noticing that the subsequent intervals
all stay below the $t^+_{LPP} $ curve.
}
 \label{F.Hierarchygamma}
 \end{figure}

Apart from the thresholds in the depolarizing rates, we can also
compare the time windows in which the ${\cal N}$--rates are
positive and the non-Markovian nature of the dynamics explicitly
appears. These are the time intervals effectively contributing to
the integral of Eq.~\eqref{E.definitionsnm}. For the first three
measures these can be cast into the form $\tan{t}^+_m\in[
y^-_m, y^+_m]$, 
where
$y^\pm_m{=}({b_m{\pm}\sqrt{b^{2}_m{-}4z^2}})/{2z^2}$, $m$ is a label
identifying the various measures, and
\begin{equation}
\label{times} b_{RHP}=2b_{BLP}{=}3b_{LPP}{=}{2(z^2-{}1)}/{\gamma}.
\end{equation}
Differently from the other ${\cal N}$--rates, $X_{LFS}$ does not
allow for an analytical treatment, and both the threshold for
Markovianity onset and the time window have been evaluated
numerically, see Fig. (\ref{F.Hierarchygamma}). It is interesting to notice that for the $N_{LFS}$ measure there
is a {\it critical value} of the rate $\gamma$ beyond which the
evolution of $\sys$ is Markovian regardless of the value of $z$.
This feature is unique of such measure, as the remaining ones
allow for values of $(\gamma,z)$ such that the dynamics remains
non-Markovian for any $\gamma$.

The results presented so far are fully consistent with what is
found by using a time-local master equation for the reduced
dynamics of $\mathcal{S}$. By following the approach outlined in
Ref.~\cite{AnderssonCH07}, this reads
\begin{equation}
\label{ME}
\dot{\rho}^{\sys}_t{=}{-}i[{\lambda\hat\sigma^{\sys}_z},{\rho}^{\sys}_t]{+}\!\!\!
\sum_{i{=}\pm,z}\!\!\Gamma_i(t)\!\left([\hat\sigma^{\sys}_i,\rho_t
{{\hat\sigma}^{{\sys}\dagger}_i}]{+}[{\hat\sigma}^{\sys}_i\rho_t,
{{\hat\sigma}^{{\sys}\dagger}_i}]\right),
\end{equation}
where we introduced the ladder operators
$\hat\sigma^{\sys}_{\pm}{=}(\hat\sigma^{\sys}_x{\pm}i\hat\sigma^{\sys}_y)/2$,
the parameters $\Gamma_\pm(t){=}\gamma$ and
$\Gamma_z(t){=}\frac{1}{2}[\gamma{-}\frac{(z^2{+}1)\sin(2
t)}{1-(1-z^2)\sin^2 t}]$, and the frequency shift $\lambda$ for
the system.
The map generated by Eq.~\eqref{ME} is non-Markovian when
$\Gamma_z(t){<}0$, which yields the very same threshold as
$\gamma^*_{RHP}$.

We have thus found a hierarchy of non-Markovianity in the
model under study which orders the four measures that we have
addressed using the depolarizing-rate thresholds
\begin{equation}
\gamma_{RHP}^*\geq\gamma_{BLP}^*\geq\gamma_{LPP}^*\geq\gamma_{LFS}^*.
\end{equation}
The equalities hold only for $z{=}1$ [cf.
Fig.~\ref{F.Hierarchygamma} {\bf (a)}]. Moreover, as shown in Fig.~\ref{F.Hierarchygamma} {\bf (b)}, we have
identified a nested structure of the time windows contributing to
the integral measures in Eq.~\eqref{E.definitionsnm}, showing that
\begin{equation}
t^+_{LFS}{\subset}t^+_{LPP} {\subset}t^+_{BLP} {\subset} t^+_{RHP}
\label{eqtempi} .
\end{equation}
While highlighting in a clear and original way the inherent
differences in the various measures that we have considered, each
addressing a different facet of non-Markovianity, this result
paves the way to the analysis of other dynamical models, in an
attempt to establish a universal, model-independent hierarchy
\cite{supplenota}. Moreover, all of the measures discussed here
share the same extremal behavior w.r.t. the initial population of
the non-Markovian environment $\mathcal{E}_{nM}$, \textit{i.e.},
at $z{=}1$ all measures give zero, whereas at $z{=}0$ they achieve
their maximum values. Remarkably, at this point, the dynamics of
$\mathcal{S}$ cannot be made Markovian by adding isotropic
depolarizing noise for any value of the decay rate according to
three of the measures here employed. On the contrary, the LFS
measure gives the threshold $\gamma^*\sim0.6$ for the onset of
Markovian dynamics. Finally, a further common feature to all the
measures here addressed is that the non-Markovian behavior of the
open system is enhanced by increasing the degree of mixedness of
$\mathcal{E}_{nM}$.

In order to establish a link between the features discussed here
and a situation of experimental relevance, we discuss the case of
a single electron quantum dot (QD)~\cite{HansonKPTV07,
Urbaszeketal13}, which, {\textit{inter alia}}, constitutes one of
the most promising platforms for quantum information
processing~\cite{ima,KloeffelL13}.
While the electron  constitutes the open system $\mathcal{S}$, the
nuclear spins surrounding it might embody an instance of
$\mathcal{E}_{nM}$. Moreover, stray phonon excitations due to
impurities in the substrate onto which the dot has grown provide
an environment that is large enough and sufficiently weakly
coupled to $\sys$ to be responsible for a Markovian channel
\cite{HansonKPTV07}. In the following, to fix the ideas,  we will refer
to the case of a semiconductor quantum dot of the II-VI group, such as, e.g., CdTe/ZnTe or Cd/Se QDs~\cite{Wu10, LeGalletal12, Guptaetal99}.
However, our results are fully general, do not depend on the specific instance
of physical system at hand, and can be applied to other similar
physical situations which are described by the central-spin model, such as the
NV$^-$-center at large magnetic fields (Supplementary Material of Ref.~\cite{Hansonetal08}),
and a single electron spin phosphorus donors in isotopically enriched ${}^{28}$Si crystal
with a reduced abundance of ${}^{29}$Si \cite{Tyryetal12, WitzelDS06}.

%

The interaction of the electron spin loaded into the quantum dot
with the nuclear ones is described by the Fermi contact hyperfine
Hamiltonian $\hat{\cal H}_{\mathcal{S},\mathcal{E}^{nM}}{=}
\sum_{n{=}1}^N J_n \hat{\bm{\sigma}}{\cdot}\hat{\bm{I}}^n$, where
$I{=}\frac{1}{2}$ for the $N{\simeq}10^3$ nuclei in the QD. The
coupling strength is proportional to the electronic envelop
wavefunction $\left|\psi\left(\bm{r}_n\right)\right|^2$ at the
$n^{\rm th}$ nuclear position.
Without loss of generality for our purposes, we use the so-called
box model for the electronic wave function
$\left|\psi\left(\vec{r}_n\right)\right|^2{=}{1}/{V}$ (with $V$
the volume occupied by the QD), which implies homogeneous
electron-nucleus couplings. We assume that a strong magnetic field
is applied to the QD so that we can legitimately retain the sole
longitudinal coupling. By introducing the collective operator
$\hat{\bm {\mathcal{I}}}{=}\sum_{n{=}1}^N \hat{\bm{I}}^n$ for the
nuclear spins and invoking the rotating-wave approximation, the
interaction Hamiltonian takes an Ising-like form $\hat{\cal H}{=}
J \hat{\sigma}^z\hat{\mathcal{I}^z}$~\cite{HansonKPTV07,
Tayloretal07}. In the mean-field approach, the fluctuations of the
nuclear magnetic field along the $z$-axis, commonly known as
Overhauser field~\cite{Urbaszeketal13}, are responsible for the
dephasing of the electron-spin state~\cite{KhaetskiiLG02}. The
typical timescale for such an effect is $T_2\simeq 2.5 \unit{ns}$
at zero magnetic field~\cite{Guptaetal99} as a result of the
averaging over many Overhauser field configurations. However, in
the presence of strong external magnetic fields and with the
application of dynamical-decoupling techniques,   coherence times
$T_2$ up to seconds have have been achieved
experimentally~\cite{Awschalometal13}. In the following, we assume
such conditions of negligible nuclear-spin-induced dephasing. As
an additional remark, we notice that dissipation induced by the
relaxation of nuclear spins due to the dipole-dipole interaction
(which is not total angular momentum conserving)  occurs at much
longer timescales $T_1$~\cite{MerkulovER02}.

Within these assumptions, the only environmental effects on the
electron spin  should be ascribed to fluctuating electrical fields
caused by or by technical noise (fluctuating gate potentials,
spurious background charges or fabrication defects). In the
following we show that, even if these noise sources could be
modelled by Markovian mechanisms, the dynamics of the electron
will never
be Markovian.

The dynamics of the open system coupled to $N$ mutually
non-interacting spins as described above, under the assumption of
a general environmental action characterized by the depolarising
parameter $\gamma$, yields the same map
$\rho^{\mathcal{S}}_t{=}\phi_t\rho^{\mathcal{S}}_0$ reported
above. Here, however, we have $G_t{=}e^{-\gamma t} {\rm Tr} [e^{-
i \hat {\mathcal{I}}^z t} \rho^{\enm}]$, where $\hat
{\mathcal{I}}^z{=}\sum_{n=1}^N \hat I^z_n$ is the collective
nuclear spin operator along the $z$-axis, $\rho^{\enm}$ is the
density matrix of the $N$ spins that make up $\mathcal{E}_{nM}$
and we took again $J$ as a frequency unit. Although this
expression can be evaluated for any environmental spin state, here
we restrict our attention to the experimentally motivated case of
$\rho^{\enm} {=}\otimes^N_{p=1}\rho_{nM}^p$. In this case, the
single-spin state $\rho_{{nM}}^p$ is the same regardless of $p$.
These assumptions entail
\begin{equation}
\begin{aligned}
{\rm Tr}[e^{- i \hat {\mathcal{I}}^z t}\rho^{\enm}]&{=}\!
\sum_{k{=}0}^N\!\binom {N} {k}\!
\frac{e^{-i I^z t \left(2k-N\right)}}{2^N}({1\!+\!z})^k ({1\!-\!z})^{N-k}\\
&{=}\left[\cos t +i z \sin t\right]^N,
\end{aligned}
\end{equation}
where $k$ is the number of environmental spins in the $\ket{0}$
state. A straightforward calculation shows that the threshold for
the onset of Markovian dynamics increases linearly with $N$ as
$\gamma_{{N}_{RHP}}^*{=}N \gamma_{RHP}^*$, whereas the time
intervals at which $\partial_t{X}{>}0$ in
Eq.~\eqref{E.definitionsnm} shrinks due to $b_{RHP}{=}2 N
\left(z^2{-}1\right)/\gamma$. Considering that, within the range
of temperatures typical for quantum dots,
$\rho^{\enm}{=}2^{-N}{\openone}$, {\textit{i.e.}}, $z{=}0$, we
obtain the remarkable result that the open system dynamics of the
dot \textit{cannot} be made Markovian by adding a Markovian noise
of whatever rate. In addition, for $N{\rightarrow}\infty$, we
obtain $\mathcal{N}_{RHP}{=}1$,
$\mathcal{N}_{BLP}{=}(e^{\pi\gamma}-1)^{-1}$ and
$\mathcal{N}_{LPP}=(e^{3\pi\gamma}-1)^{-1}$, meaning that the
dynamics of $\sys$ is non-Markovian for any initial state of
$\mathcal{E}_{nM}$ that is not an eigenstate of $\hat
{\mathcal{I}}^z$. As seen above, for $z{=}1$ the reduced dynamics
is Markovian regardless of $\gamma$ and $N$.

We have investigated the behavior of the open-system dynamics of a
spin-$\frac{1}{2}$ subject to the competing action of Markovian
and non-Markovian environments, identifying the conditions under
which the system evolution becomes Markovian. These account in
well defined thresholds in the the relative strength of the
coupling of the system to the various environments. A nested
hierarchical structure then results for the measures of
non-Markovianity that we have considered, which suggests a
causality relation amongst the different physical phenomena used
to characterize non-Markovianity in this context. Our findings
might be used to acquire information on the open-system dynamics
of a single-electron QD for which, under fairly reasonable
assumptions, a Markovian description of the dynamics turns out to
be fully inadequate.

\noindent {\it Acknowledgments.--} TJGA and CDF thank L.~
Mazzola and A.~ Xuereb for useful discussions. TJGA is supported
by the European Commission, the European Social Fund and the
Region Calabria through the program POR Calabria FSE
2007-2013-Asse IV Capitale Umano-Obiettivo Operativo M2. MP thanks
the UK EPSRC (grant nr. EP/G004579/1), the Alexander von
Humboldt Stiftung, and the John Templeton Foundation (grant ID 43467) for financial support.




\section*{Appendix: analysis of other decoherence channels}

In this Appendix we extend the analysis presented in the main text
to the case of a more general decoherence channel given by a
combination of an amplitude damping and a depolarizing ones. The resulting channel is
characterized by the Kossakowski matrix
\begin{equation}
  {\bm \gamma}{=}
  \begin{pmatrix}
\gamma_A & i\gamma_A/2 & 0 \\
{-}i\gamma_A/2 & \gamma_A & 0 \\
0 & 0 & 0
\end{pmatrix}
+
\begin{pmatrix}
\gamma_D  & 0 & 0 \\
0 & \gamma_D & 0 \\
0 & 0 & \gamma_D
\end{pmatrix}.
\end{equation}
The case $\gamma_A=0$ has been already analized in the main text. In the presence
of the amplitude damping component, the explicit expressions for
the ${\cal N}$-rates are analogous to the ones reported in the
main text in Eq. (\ref{analitiche}). By introducing $f_t^{\pm}=1\pm e^{-4(\gamma_A+\gamma_D)t}$ 
we have
\begin{equation}
\label{analitiche2}
\begin{aligned}
X_{BLP}&=\partial_t|G_t|,\\
X_{LPP}&=\partial_t[(f^+_t{-}1) |G_t|^2],\\
X_{RHP}&=\lim_{\tau \to 0^+}\frac{1}{2\tau}\left(g(t,\tau){-}\frac{f^+_\tau}{2}{+}\left|g(t,\tau){-}\frac{f^+_\tau}{2}\right|\right),\\%
X_{LFS}&{=}
\partial_t{\cal I}(\sigma),
\end{aligned}
\end{equation}
where $\mathcal{I}(\sigma )$ stands for the Quantum mutual
information of the matrix
\begin{equation}
\sigma =
\frac{1}{4}
\begin{pmatrix}
 f^+_t{-}\frac{\Upsilon}{2} f^-_t & 0 & 0 & 2 G^*_t \\
 0 & f^-_t{-}\frac{\Upsilon}{2} f^-_t  & 0 & 0 \\
 0 & 0 & f^-_t{+}\frac{\Upsilon}{2} f^-_t  & 0 \\
 2 G_t & 0 & 0 & f^+_t{+}\frac{\Upsilon}{2}f^-_t \\
\end{pmatrix},
\end{equation}
where we have introduced the parameters
\begin{equation}
\begin{aligned}
&\Upsilon=\dfrac{\gamma_A}{(\gamma_A+\gamma_D)},~~G_t=\sum_{p{=}\pm}e^{-2(\gamma_A+2\gamma_D)
t{+} i p t}B^{pp}_0,\\
&g(t,\tau)=\sqrt{\left(\frac{\beta}{4}f^-_{\tau}\right)^2+\left
| \frac{G_{t+\tau}}{G_t}\right |^2}.
\end{aligned}
\end{equation}
The dynamics of $\sys$ turns out to be non-Markovian for any triplet $\gamma_A$, $\gamma_D$, $z$ such that the quantities in
Eqs.~(\ref{analitiche}) are positive. For the ${\cal N}_{RHP}$ measure, this condition leads to the
inequality
\begin{equation}
4\gamma_D{-}\frac{(z^2-1)\sin(2 t)}{\cos^2 t{+}z^2 \sin^2 t}{<}0.
\label{ineqRHP2}\end{equation} As this function does not depend on
$\gamma_A$, this implies that a Markovian behavior occurs for
decoherence rates satisfying the same threshold valid for the sole
depolarizing channel.

\begin{figure}[t]
{\includegraphics[width=.7\columnwidth]{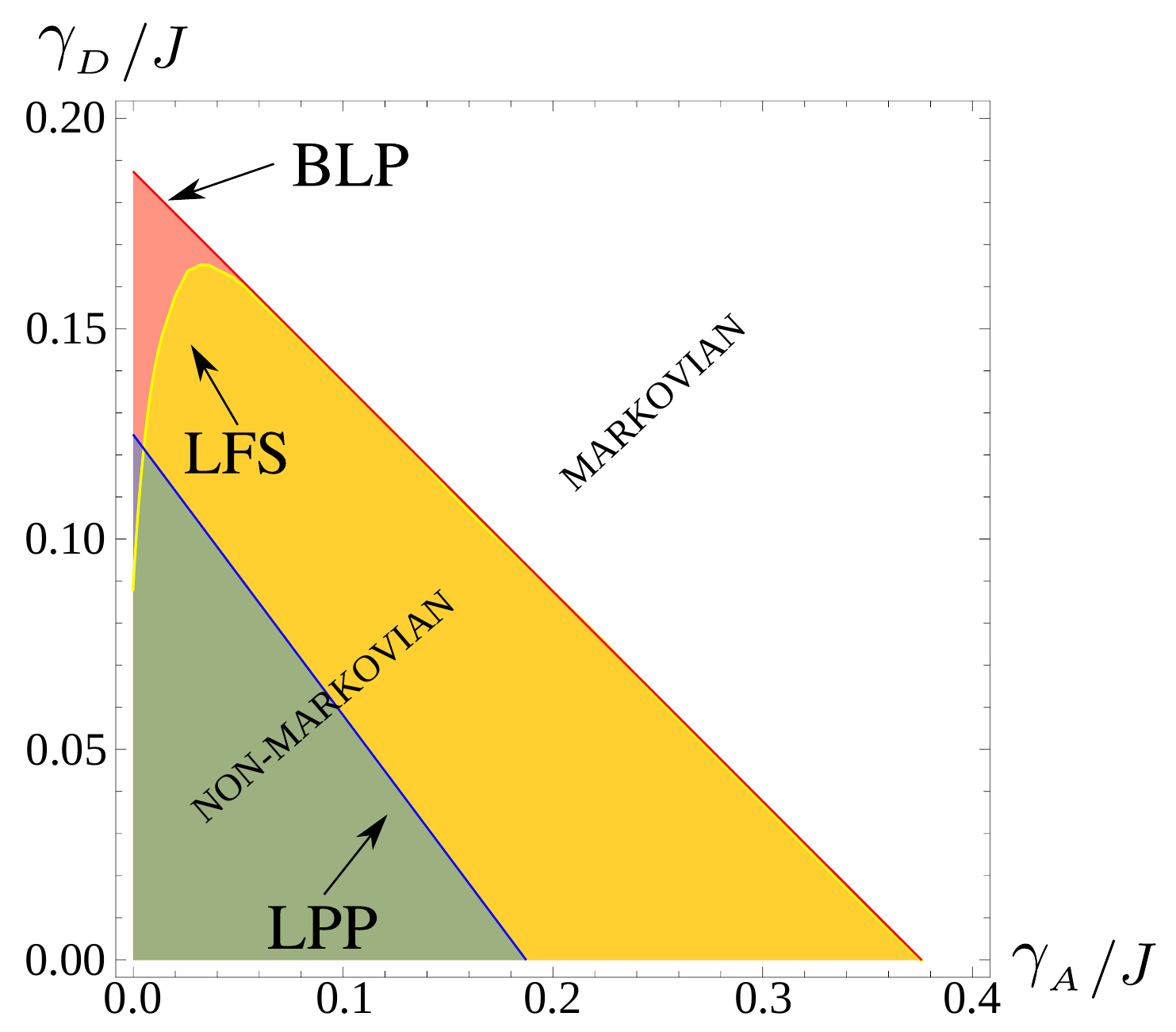}}
\caption{(Color online) Markovianity phase diagram in the
$(\gamma_A,\gamma_D)$ plane, with $z=0.5$ (other values of the
initial magnetization for ${\cal E}_{nM}$ give rise to similar
behaviors). The dynamics of the system is Markovian within the
colored regions according to the various non-Markovianity
measures. The borders of these regions (pointed by the arrows)
represent the threshold conditions discussed in the text. The LFS
measure has a different behavior for small $\gamma_A$, while it
tends to coincide with the others quantifiers when the rate of the
amplitude damping channel gets larger. }
\label{gAvsgD}
 \end{figure}

For the ${\cal N}_{BLP}$ and the $N_{LPP}$  measures, on the other
hand, the positivity condition leads to
\begin{equation}
\label{ineqLPPBLP}
4(\gamma_A+m\gamma_D)+\frac{(1-z^2)\sin(2Jt)}{\cos^2(Jt)+z^2
\sin^2 (Jt)}{<}0,
\end{equation}
with $m=2$ ($m=3$) for the BLP (LPP) measure. 
For the BLP case, the onset of Markovian dynamics is found to
occur for $\gamma_A+2\gamma_D{=}(1{-}z^2)/{2 z}$, while the
analogous condition for the LPP measure reads
$2\gamma_A+3\gamma_D{=}(1{-}z^2)/{3 z}$. The form taken by the LFS measure does not allow for
an analytic expression. Its behavior is shown in Fig.~\ref{gAvsgD} as a function
of $\gamma_A$ and $\gamma_D$, at a set value of $z$, and is compared to the BLP and LPP measures so as to establish a Markovianity phase-diagram valid for the respective measure.

A hierarchical relation among the time windows that contribute to the evaluation of the different indicators can be
found here too and leads precisely to the structure given in 
Eq. (\ref{eqtempi}).

\end{document}